\documentclass{article}
\PassOptionsToPackage{numbers, compress}{natbib}
\usepackage[final]{neurips_2023_ml4ps}
\usepackage[utf8]{inputenc} % allow utf-8 input
\usepackage[T1]{fontenc}    % use 8-bit T1 fonts
\usepackage{hyperref}       % hyperlinks
\usepackage{url}            % simple URL typesetting
\usepackage{booktabs}       % professional-quality tables
\usepackage{amsfonts}       % blackboard math symbols
\usepackage{nicefrac}       % compact symbols for 1/2, etc.
\usepackage{microtype}      % microtypography
\usepackage{xcolor}         % colors
\usepackage{graphicx}
\definecolor{seagreen}{rgb}{0.190, 0.525, 0.361}
\definecolor{midnightBlue}{rgb}{0.098, 0.098, 0.439}
\definecolor{darksalmon}{rgb}{0.914, 0.588, 0.478}
\definecolor{bottlegreen}{rgb}{0.08, 0.45, 0.1}
\definecolor{anthracite}{rgb}{0.34, 0.38, 0.42}
\definecolor{royalpurple}{rgb}{0.47, 0.317, 0.66}%\definecolor{anthracite}{rgb}{0.271, 0.270, 0.318}
\title{The search for the lost attractor}
\author{
Mario Pasquato\\
\small{Département de Physique, Université de Montréal, Montréeal, Quéebec, Canada}\\
\small{Ciela Institute, Montr\'eal, Qu\'ebec, Canada}\\
\small{Mila - Québec Artificial Intelligence Institute, Montréal, Québec, Canada}\\
\small{Dipartimento di Fisica e Astronomia, Università di Padova, Padova, Italy}\\
\small{\texttt{mario.pasquato@umontreal.ca}}\\
\vspace{-0.3in}
\And
Syphax Haddad\\
\small{Département de Physique, Université de Montréal, Montreal, Quebec, Canada}\\
\small{Cégep Trois Rivières, Trois Rivières, Quebec, Canada}\\ %Pavillon des Sciences, 3500 Rue de Courval, 
\vspace{-0.3in}
\And
Pierfrancesco Di Cintio\\
\small{National Council of Research - Institute of Complex Systems, Sesto Fiorentino, Italy}\\
\small{INAF - Osservatorio Astrofisico di Arcetri, Firenze, Italy}\\
\small{INFN - Sezione di Firenze, Sesto Fiorentino, Italy}\\
\vspace{-0.3in}
\And
Alexandre Adam\\
\small{Département de Physique, Université de Montréal, Montréal, Québec, Canada}\\
\small{Ciela Institute, Montr\'eal, Qu\'ebec, Canada}\\
\small{Mila - Québec Artificial Intelligence Institute, Montréal, Québec, Canada}\\
\vspace{-0.3in}
\And
Pablo Lemos\\
\small{Département de Physique, Université de Montréal, Montréal, Québec, Canada}\\
\small{Ciela Institute, Montr\'eal, Qu\'ebec, Canada}\\
\small{Mila - Québec Artificial Intelligence Institute, Montréal, Québec, Canada}\\
\small{Center for Computational Astrophysics, Flatiron Institute, NYC, NY, USA}\\ %162 5th avenue, NYC,
\vspace{-0.3in}
\And
Noé Dia\\
\small{Département de Physique, Université de Montréal, Montréal, Québec, Canada}\\
\small{Ciela Institute, Montr\'eal, Qu\'ebec, Canada}\\
\small{Mila - Québec Artificial Intelligence Institute, Montréal, Québec, Canada}\\
\vspace{-0.3in}
\And
Mircea Petrache\\
\small{Facultad de Matemáticas \& Instituto de Ingeneria Matematica y Computacional, Santiago, Chile}\\
\vspace{-0.3in}
\And
Ugo Niccolò Di Carlo\\
\small{SISSA - Scuola Internazionale Superiore di Studi Avanzati, Trieste, Italy}\\
\small{INFN - Sezione di Trieste, Trieste, Italy}\\
\vspace{-0.3in}
\And
Alessandro Alberto Trani\\
\small{Niels Bohr Institute, Copenhagen, Denmark}\\ %Blegdamsvej 17, DK-2100 
\small{Research Center for the Early Universe, School of Science, The University of Tokyo, Tokyo, Japan}\\ %tel:113-0033
\small{Okinawa Institute of Science and Technology, Okinawa, Japan}\\ %1919-1 Tancha, Onna-son, tel:904-0495,
\vspace{-0.3in}
\And
Laurence Perreault-Levasseur\\
\small{Département de Physique, Université de Montréal, Montréal, Québec, Canada}\\
\small{Ciela Institute, Montr\'eal, Qu\'ebec, Canada}\\
\small{Mila - Québec Artificial Intelligence Institute, Montréal, Québec, Canada}\\
\small{Center for Computational Astrophysics, Flatiron Institute, NYC, NY, USA}\\ %162 5th avenue,
\vspace{-0.3in}
\And
Yashar Hezaveh\\
\small{Département de Physique, Université de Montréal, Montréal, Québec, Canada}\\
\small{Ciela Institute, Montr\'eal, Qu\'ebec, Canada}\\
\small{Mila - Québec Artificial Intelligence Institute, Montréal, Québec, Canada}\\
\small{Center for Computational Astrophysics, Flatiron Institute, NYC, NY, USA}\\ %162 5th avenue, 
}
\begin{document}
\maketitle
\begin{abstract}

N-body systems characterized by $r^{-2}$ attractive forces may display a self-similar collapse known as the gravo-thermal catastrophe or core-collapse. Globular clusters are a real-life example of this in astronomy. In these clusters, collapse is halted by dynamical heating from binary stars. This is known as the binary-burning phase of star cluster evolution. A fraction of Milky Way globular clusters may have already reached this phase.
It has been speculated --with guidance from simulations-- that macroscopic variables such as central density and velocity dispersion are governed post-core-collapse by an effective, low-dimensional system of ordinary differential equations (ODEs). However, it is hard to distinguish potentially chaotic low-dimensional motion, from high-dimensional stochastic noise. Here we apply three machine learning tools to the time series of relevant macroscopic quantities from state-of-the-art dynamical simulations to constrain the post-collapse dynamics: topological data analysis (TDA) on a lag embedding, Sparse Identification of Nonlinear Dynamics (SINDY), and Tests of Accuracy with Random Points (TARP). Even though TARP suggests that the time series are not purely noise, we do not find a low-dimensional system of equations describing their time evolution.
\end{abstract}
%%%%%%%%%%%%%%%%%%%%%%%
\section{Introduction}
Point particles interacting through inverse square gravitational forces are ubiquitous in astronomy: from galaxies in galaxy clusters, to stars in galaxies and star clusters, down in scale to planetesimals and even dust \citep[][]{Trenti:2008}. While the general gravitational N-body problem is approached numerically either by brute forcing the pairwise interactions \citep[direct N-body;][]{2003gnbs.book.....A} or by approximated schemes (multi-particle collision, MPC \cite{2021A&A...649A..24D}, Monte-Carlo \cite{2000MNRAS.317..581G}), it is expected that in specific regimes there may emerge tractable effective equations driving the evolution of relevant macroscopic quantities. It has been suggested that this is the case of post core-collapse oscillations in star clusters \cite{1995ApJ...448..672B, 2013MNRAS.432.2779B}. E.g. \cite{1995ApJ...448..672B} claim to find a three-dimensional attractor that qualitatively resembles the R\"ossler attractor \citep[see][]{Letellier:2006}. 

\textbf{Paper contribution:} For the first time, we apply three independent machine learning tools to analyze state-of-the-art simulations of star cluster core collapse. We do not find clear evidence of effective low dimensional dynamics. Topological data analysis \citep[TDA; ][]{wasserman2018topological} when applied to a lag embedding of the core density and velocity dispersion time series produces a persistence diagram that appears indistinguishable from noise; similarly, Sparse Identification of Nonlinear Dynamical Systems \citep[SINDY; ][]{brunton2016sparse} fails to yield an equation reproducing the observed time series. Finally, Tests of Accuracy with Random Points \citep[TARP; ][]{2023arXiv230203026L} find that samples from our original time series do not lie in-distribution with respect to random reshufflings. The post core-collapse evolution is thus not purely independent identically distributed noise, but this is not sufficient to conclude that it is the result of low-dimensional dynamics.

\section{Simulations}
\begin{figure}
  \centering
    \begin{minipage}{0.5\linewidth}
        \includegraphics[width=\linewidth]{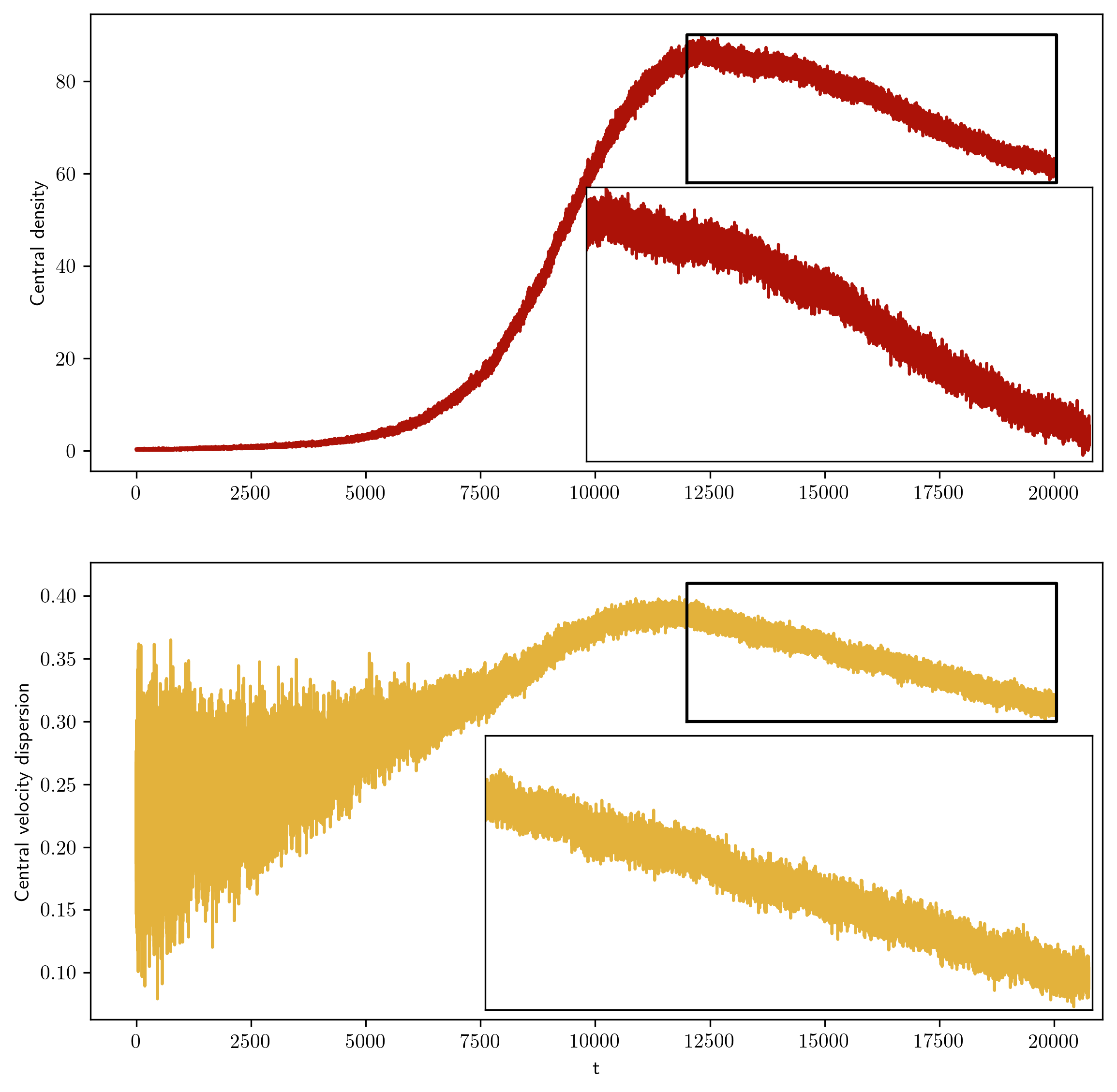}
    \end{minipage}%
    \begin{minipage}{0.5\linewidth}
        \includegraphics[width=\linewidth]{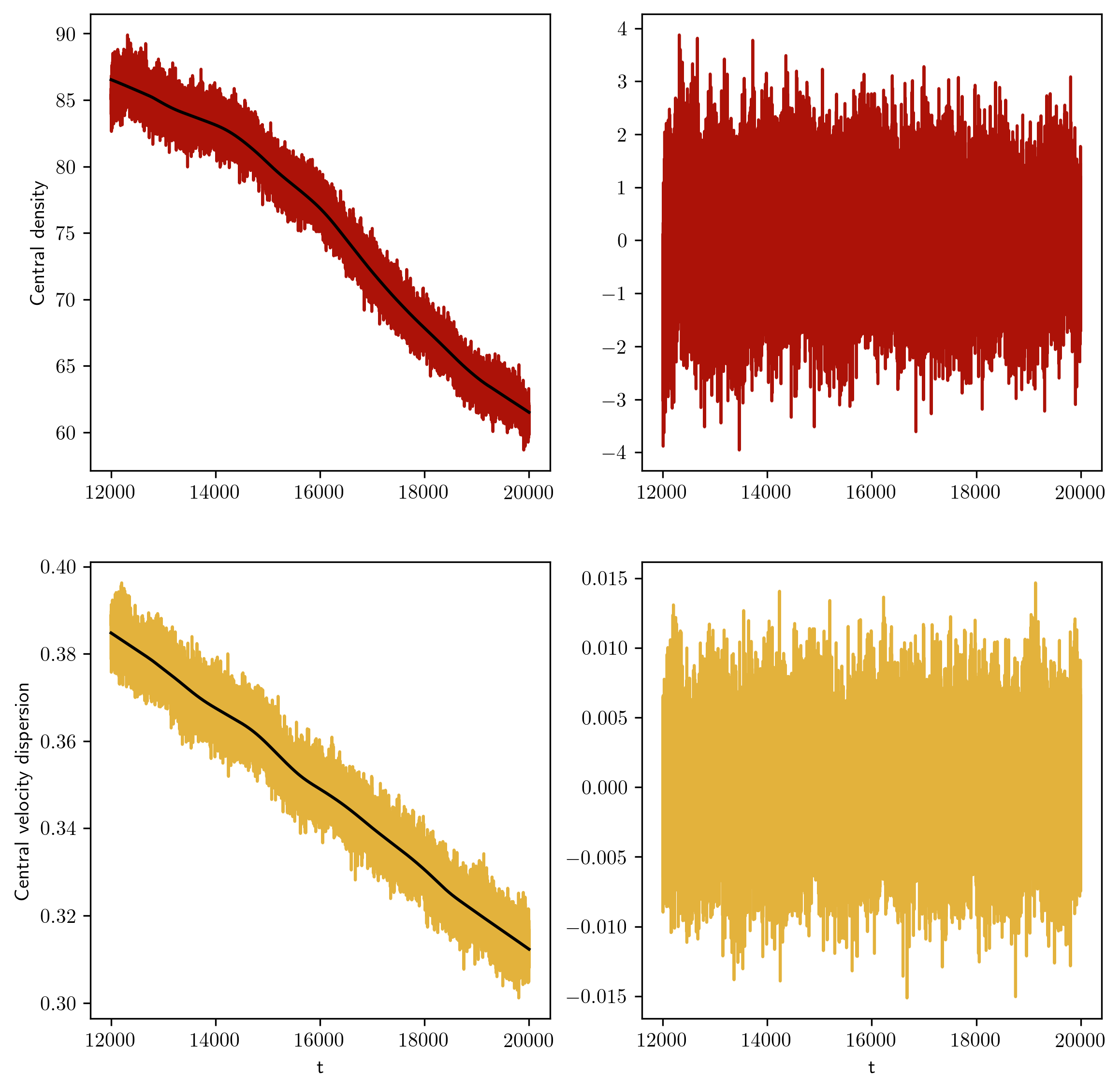}
    \end{minipage}%
  \caption{\textbf{Left panel:} Evolution of density (top, red) and velocity dispersion (bottom, yellow) in our simulation. Zoom in on the post-core-collapse part of the time series (inset panels). \textbf{Right panel:} Trend fitting (left, black solid lines) and detrended time series (right).\label{rhosigma}}
\end{figure}

We have performed a hybrid MPC-particle-mesh simulation with $N=10^5$ particles where the collective gravitational force is evaluated with a standard particle in cell method on a $32\times 16\times 128$ spherical grid. The Poisson equation is solved with a finite difference scheme. Collisions are resolved with the MPC stochastic scheme \cite{2021A&A...649A..24D} on the same grid, conditioned every time step $\Delta t$ with the cell-based interaction probability $
p_i={\rm Erf}\left(\beta\Delta t \nu_c\right)$,
where $\nu_c={8\pi G^2\bar{m}^2_i n_i\log\Lambda}/{\sigma^3_i}$ is the typical collision frequency in cell $i$ and $\beta$ a dimensionless constant proportional to the total number of cells.
Initial positions and velocities are sampled from an isotropic Plummer model with scale radius $r_s$ and mass scale $M$.
The equations of motion for the $N$ particles are propagated between each collision step with a standard second order leapfrog scheme with a fixed timestep. All radii are expressed in  units of the scale radius $r_s$, while the time unit is given as a function of the total mass $M$ and the gravitational constant $G$ as $t_{\rm dyn}\equiv\sqrt{r_s^3/GM}$. We fix $\Delta t=t_{\rm dyn}/100$ and run until $2\times 10^5t_{\rm dyn}$. As a function of time, we save the values of the central density and velocity dispersion (both evaluated within the time-dependent radius enclosing the $5 \%$ of M), as shown in Fig.~\ref{rhosigma}, left panel.

\section{Methods}
\subsection{Topological data analysis on lag embedding}
TDA is the study of the topological properties of point clouds derived from data. The primary tool within TDA is persistent homology, which provides a multi-scale description of the homological features of a dataset such as connected components ($H_0$), loops ($H_1$), and voids ($H_2$). The main output of persistent homology is the barcode or persistence diagram (PD), visualizing the birth and death of topological features as one varies a scale parameter. See \cite{chazal2021introduction, munch2017user, chazal2016structure, dey2022computational} for comprehensive discussion.

We applied persistent homology to a lag embedding of the time series of velocity dispersion and of central density generated from our simulations. To achieve this, we deterended our time series by subtracting a polynomial fit (see Fig.~\ref{rhosigma}, right panel), calculated the autocorrelation and sampled at regular intervals such that the autocorrelation signal became insignificant. This was the basis for computing a four-dimensional lag embedding. Our lag embeddings reconstruct a higher-dimensional phase space from a single time series, effectively transforming a one-dimensional series into a point cloud. This is justified by Takens' theorem, asserting that, under certain conditions, the dynamics of a system can be reconstructed using the time-delayed versions of even a single observable \citep[][]{takens2006detecting}. 

By applying TDA to this lag-embedded space, we can discern topological features that might correspond to dynamical structures or patterns in the original time series. For instance, loops revealed by persistent homology might indicate periodic orbits or recurrent dynamics. We used the \emph{ripser} library \cite[][]{ctralie2018ripser} to compute persistent homology features on our timeseries.

\subsection{Sparse Identification of Nonlinear Dynamics (SINDY)}
SINDY is an algorithmic approach designed to derive governing equations of dynamical systems directly from observational data \citep[][]{brunton2016sparse}. The central idea is to compute derivatives of the observed state variables and then express these derivatives as a sparse linear combination of a pre-defined library of functions. Here we relied on a polynomial library up to third degree, with interactions. We filtered our time series by removing high-frequency Fourier modes and took first order numerical derivatives of velocity dispersion and of central density generated from our simulations. We thus used SINDY with LASSO regularization and MSE loss to look for second-order ODEs, the minimum order that allows for oscillatory behavior. At the implementation level, we relied on the \emph{pysindy} library \citep[][]{2022JOSS....7.3994K}.

\subsection{Tests of Accuracy with Random Points}
TARP is a necessary and sufficient coverage test to assess whether a set of points has been drawn from a given distribution for which a generative model is available \citep[][]{2023arXiv230203026L}. TARP has been originally designed to test the accuracy of posterior estimators by using random points to assess coverage probabilities, but here we are applying it essentially as an anomaly detection technique, for the task of distinguishing our original time series from random reshuffled versions of it. For our purposes, the main output of TARP is the expected coverage probability as a function of credible intervals. When plotted, this reveals that a sample is out of distribution if it deviates from the identity line. Unlike the previous two approaches, finding such a deviation through TARP does not prove, per se, that the dynamics producing our time series in low dimensional.

%\begin{figure}
%\centering
%\includegraphics[width=\linewidth]{density data and fft.png}
%\caption{\label{fig:graph 1} Figure showing density data before and after noise. Top left, the original density data; bottom left, the smoothed data after applying the Fourier transform (top right); bottom right, the numerical derivative of the smoothed data.}     
%\end{figure}

%\begin{figure}
%\centering
%\includegraphics[width=\linewidth]{sigma data and fft.png}
%\caption{\label{fig:graph 2} Figure showing dispersion velocity data before and after noise. Top left, the original data; bottom left, the smoothed data after applying the Fourier transform (top right); bottom right, the numerical derivative of the smoothed data.}     
%\end{figure}

%\begin{figure}
%    \centering
%    \includegraphics[width=\linewidth]{sim rho and sigma without t - lamda 8.1728e-6.png}
%    \caption{Figure showing the simulation of our SINDy model (red dotted curves) on the data after the Fourier transform of the central density and dispersion velocity as well as their numerical derivatives (black curves). The model optimization parameters are $\alpha=1e^{-5}$ and $\lambda=8.1728 \times {10}^{-6}$.}
%    \label{fig:graph 5}
%\end{figure}

\section{Results}
\subsection{Topological data analysis on lag embedding}
\begin{figure}
  \centering
  \begin{tabular}{cc}
  \includegraphics[width=0.45\textwidth]{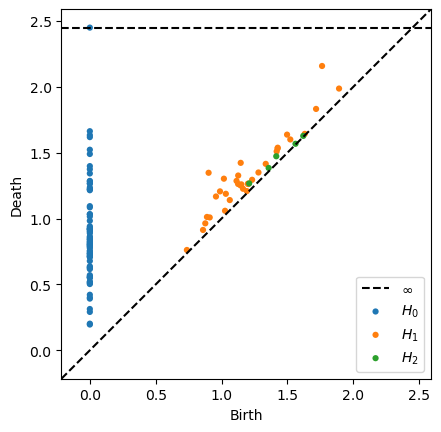} &
  \includegraphics[width=0.465\textwidth]{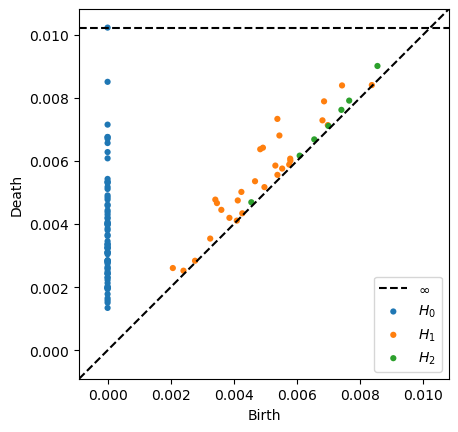}\\
  \end{tabular}
  \caption{Persistence diagrams of central density (left) and velocity dispersion (right). Blue dots ($H_0$) represent connected components, essentially subclusters of points in the point cloud. Orange dots ($H_1$) represent rings, the kind of feature we expect if the evolution in lag embedding space is (quasi) periodic, e.g. approaching a limit cycle. Green dots ($H_2$) represent voids, which may correspond to a more complex but still low-dimensional dynamics on a 2D surface.\label{pd}}
\end{figure}

The PDs of the central density and velocity dispersion time series are shown in Fig.~\ref{pd}. In a PD, stable topological structure is represented by points far from the diagonal. Qualitatively, these do not appear to be present in Fig.~\ref{pd} for rings and voids, suggesting that these topological features picked up by persistent homology are just noise. Quantitatively, we tested this by computing the persistent entropy statistic on our original time series and on 100 random re-shufflings thereof. Persistent entropy is the Shannon entropy of the lifetime of features in a persistence diagram. The persistent entropy for the original time series typically falls within the distribution of the reshuffled time series, both for density and velocity dispersion. This is not the case for a time series obtained from the R\"ossler attractor, which happens to have systematically lower persistent entropy than its reshufflings.
%It is shown in Fig.~\ref{pe} that the persistent entropy for the original time series typically falls within the distribution of the reshuffled time series. This is not the case for a time series obtained from the Rossler's attractor (bottom row of Fig.~\ref{pe}).

%\begin{figure}
%  \centering
% \begin{tabular}{lll}
%  \includegraphics[width=0.32\textwidth]{pe_rho0.png} & \includegraphics[width=0.32\textwidth]{pe_rho1.png} & \includegraphics[width=0.32\textwidth]{pe_rho2.png} \\
%  \includegraphics[width=0.32\textwidth]{pe_sigma0.png} & \includegraphics[width=0.32\textwidth]{pe_sigma1.png} & \includegraphics[width=0.32\textwidth]{pe_sigma2.png} \\
%  \includegraphics[width=0.32\textwidth]{pe_rossler0.png} & \includegraphics[width=0.32\textwidth]{pe_rossler1.png} & \includegraphics[width=0.32\textwidth]{pe_rossler2.png} \\
%  \end{tabular}
%  \caption{Distribution of persistent entropy for the PD obtained by randomly reshuffling the density (lavender histogram, top row), velocity dispersion (middle) and a time series generated from the R\"ossler attractor (bottom). Columns correspond to $H_0$, $H_1$ and $H_2$ topological features. The vertical bars correspond to the persistent entropy of the actual time series.\label{pe}}
%\end{figure}

\subsection{Sparse Identification of Nonlinear Dynamics (SINDY)}
\begin{figure}
    \centering
    \begin{minipage}{0.5\linewidth}
    \includegraphics[width=\linewidth]{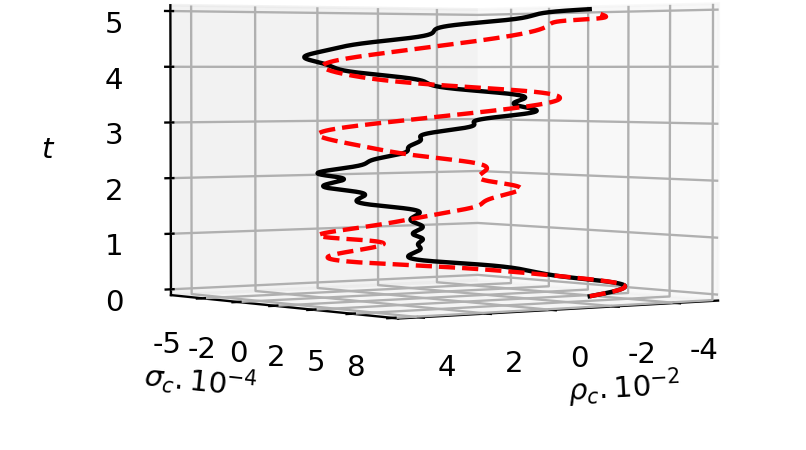}
    \end{minipage}%
    \begin{minipage}{0.5\linewidth}
    \includegraphics[width=\linewidth]{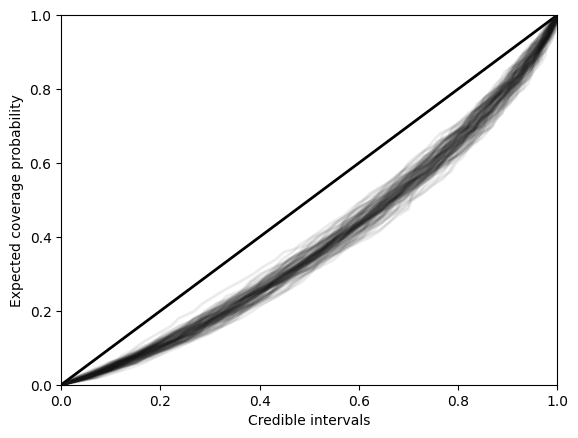}
    \end{minipage}%
    \caption{\textbf{Left panel}: solution of a hand-picked ODE found by SINDY (red dashed line) versus the smoothed time series it was learned from (solid black line). Up until $t \approx 0.5$ it follows the data closely, then it deviates catastrophically even though it still shows some oscillatory behavior. Times are in arbitrary units. \textbf{Right panel:} TARP coverage diagram. The 100 semitransparent solid curves correspond each to a run of TARP. They all present bias, systematically deviating downwards with respect to the diagonal.} %Figure showing the dispersion velocity as a function of density in 3D plot where z = t. The model optimization parameters are $\alpha=10^{-5}$ and $\lambda=8.1728 \times {10}^{-6}$.
    \label{fig:sindy}
\end{figure}

We conducted a thorough analysis of the ODEs proposed by SINDY while varying the strength of regularization, which results in different levels of sparsity in the ODE coefficients. Neither a grid search nor a meticulous manual exploration resulted in equations able to reproduce our time series in the long term. In Fig.~\ref{fig:sindy} (left panel) we show the solution of one such equation over time. While it initially reproduces the time series it is learned from, it eventually departs dramatically from it. This was the case for all systems of ODEs we found.

\subsection{Tests of Accuracy with Random Points}
Fig.~\ref{fig:sindy} (right panel) shows 100 runs of TARP, none of which overlaps the diagonal line. Our original time series is thus biased (out-of-distribution) with respect to its reshufflings, suggesting that reshuffling erases some temporal dependence. In other words, our time series is not a sequence of independent identically distributed random variables. This can be due to various reasons, including poor trend removal in the pre-processing phase; it does not \emph{per se} indicate that our time series are generated by a low dimensional dynamics. It does justify further investigation into the matter.

\section{Conclusions}
We applied three machine learning tools that are new to astronomy, especially in the context of star cluster dynamics, to the problem of characterizing the dynamical evolution of a simulated N-body system in the post core-collapse phase. TDA on a lag embedding of the central density and velocity dispersion time series, as well as SINDY, did not provide us with evidence that the dynamical evolution is driven by a low-dimensional system of ODEs, despite the expectations of \cite{1995ApJ...448..672B}.  On the other hand, TARP revealed that the time series we analyzed are out-of-distribution with respect to randomly reshuffled time series. This suggests that there is some structure in the (apparent) noise. The search for the lost attractor is not over yet.

\begin{ack}
M. P. acknowledges financial support from the European Union’s Horizon 2020 research and innovation program under the Marie Skłodowska-Curie grant agreement No. 896248. This work is supported by the Simons Collaboration on “Learning the
Universe". The Flatiron Institute is supported by the Simons Foundation. The work is in part supported by computational resources provided by Calcul Quebec and the Digital Research Alliance of Canada. Y.H. and L.P. acknowledge support from the Canada Research Chairs Program, the National Sciences and Engineering Council of Canada through grants RGPIN-2020- 05073 and 05102, and the Fonds de recherche du Québec through grants 2022-NC-301305 and 300397. P.L acknowledges support from the Simons Foundation. Mircea Petrache acknowledges financial support from Fondecyt Regular grant No. 1210426.
\end{ack}

%\section*{References}

\bibliography{neurips.bib}
\bibliographystyle{plain}
%[1] Alexander, J.A.\ \& Mozer, M.C.\ (1995) Template-based algorithms for
%connectionist rule extraction. In G.\ Tesauro, D.S.\ Touretzky and T.K.\ Leen
%(eds.), {\it Advances in Neural Information Processing Systems 7},
%pp.\ 609--616. Cambridge, MA: MIT Press.

%\section*{Supplementary material}

%\begin{figure}
%  \centering
%  \includegraphics[width=\textwidth]{autocorr.png}
%  \caption{Determining the timescale of autocorrelation.\label{autocorr}}
%\end{figure}

\end{document}